\documentclass[preprintnumbers, floatfix,twocolumn,preprintnumbers, letterpaper, superscriptaddress,nofootinbib]{revtex4}
\usepackage{amsfonts}
\usepackage{amsmath}
\usepackage{amssymb,epsf}
\usepackage{latexsym}
\usepackage{graphicx,epsfig}
\usepackage{amssymb}
\usepackage{float}
\usepackage{subfigure}
\usepackage[colorlinks=true,citecolor=blue,linkcolor=blue,urlcolor=black]{hyperref}
\usepackage{dcolumn}
\usepackage{psfrag}
\usepackage{wrapfig}
\usepackage{makeidx}
\usepackage{epsf}
\usepackage{epstopdf}

\begin{document}

\title{Analytical and numerical study of\\
backreacting one-dimensional holographic superconductors\\
in the presence of Born-Infeld electrodynamics}
\author{Mahya Mohammadi}
\affiliation{Physics Department and Biruni Observatory, Shiraz University, Shiraz 71454,
Iran}
\author{Ahmad Sheykhi}
\email{asheykhi@shirazu.ac.ir}
\affiliation{Physics Department and Biruni Observatory, Shiraz University, Shiraz 71454,
Iran}
\affiliation{Research Institute for Astronomy and Astrophysics of Maragha (RIAAM), P. O.
Box: 55134-441, Maragha, Iran}
\author{Mahdi Kord Zangeneh}
\email{mkzangeneh@scu.ac.ir}
\affiliation{Physics Department, Faculty of Science, Shahid Chamran University of Ahvaz,
Ahvaz 61357-43135, Iran}
\affiliation{Center for Research on Laser and Plasma, Shahid Chamran University of Ahvaz,
Ahvaz, Iran}

\begin{abstract}
We analytically as well as numerically study the effects of
Born-Infeld nonlinear electrodynamics on the properties of
$(1+1)$-dimensional $s$-wave holographic superconductors. We relax
the probe limit and further assume the scalar and gauge fields
affect on the background spacetime. We thus explore the effects of
backreaction on the condensation of the scalar hair. For the
analytical method, we employ the Sturm-Liouville eigenvalue
problem and for the numerical method, we employ the shooting
method. We show that these methods are powerful enough to analyze
the critical temperature and phase transition of the one
dimensional holographic superconductor. We find out that
increasing the backreaction as well as nonlinearity makes the
condensation harder to form. In addition, this one-dimensional
holographic superconductor faces with second order phase
transition and their critical exponent has the mean field value
$\beta={1}/{2}$.
\end{abstract}

\maketitle

\section{Introduction}

The most well-known theory for describing the mechanism behind
superconductivity from microscopic perspective is the BCS theory proposed by
Bardeen, Cooper and Schrieffer. According to BCS theory, the condensation of
Cooper pairs into a boson-like state, at low temperature, is responsible for
infinite conductivity in solid state system \cite{BCS57}. However, when the
temperature increases, the Cooper pairs decouples and thus the BCS theory is
unable to explain the mechanism of superconductivity for high temperature
superconductors \cite{BCS57}. The correspondence between gravity in an Anti
de-Sitter (AdS) spacetime and a Conformal Field Theory (CFT) living on the
boundary of the spacetime provides a powerful tool for calculating
correlation functions in a strongly interacting field theory using a dual
classical gravity description \cite{Maldacena}. According to the AdS/CFT
duality proposal an $n$-dimensional conformal field theory on the boundary
is equivalent to gravity theory in $(n+1)$-dimensional AdS bulk \cite%
{Maldacena,G98,W98,H08,HR08,R10}. The dictionary of AdS/CFT duality implies
that each quantity in the bulk has a dual on the boundary. For example,
energy-momentum tensor $T_{\mu \nu }$ on the boundary corresponds to the
bulk metric $g_{\mu \nu }$\cite{G98,W98}. Based on this duality, Hartnoll et
al. proposed a model for holographic superconductor in $2008$ \cite{H08}.
Their motivation was to shed light on the problem of high temperature
superconductors. According to the holographic superconductors, we need a
hairy black hole in gravity side to describe a superconductor on its
boundary. During the past decade, the investigation on the holographic
superconductor has got a lot of attentions (see e.g. \cite%
{HR08,R10,H09,Hg09,H11,Gu09,HHH08,JCH10,SSh16,SH16,cai15,Ge10,SHsh(17),CAI11,
SHSH(16), shSh(16),Doa, Afsoon,
cai10,cai14,yao13,n3,n4,n5,n6,Gan1}).

On the other hand, BTZ (Bandos-Teitelboim-Zanelli) black holes, the
well-known solutions of general relativity in $(2+1)$-dimensional spacetime,
provide a simplified model to investigate some conceptual issues in black
hole thermodynamics, quantum gravity, string theory, gauge theory and
AdS/CFT correspondance \cite{Car1,Ash,Sar,Wit1,Car2}. It has been shown that
the quasinormal modes in this spacetime coincide with the poles of the
correlation function in the dual CFT. This gives quantitative evidence for
AdS/CFT \cite{Bir}. In addition, BTZ black holes play a crucial role for
improving our perception of gravitational interaction in low dimensional
spacetimes \cite{Wit2}. These kind solutions have been studied from
different point of views \cite{rin1,rin2,rin3,rin4}.

Holographic superconductors dual to asymptotic BTZ black holes
have been explored widely (see e.g. \cite{Wang,chaturvedi,L12,momeni,peng17,lashkari,hua,yanyan,yan,alkac,50-1,bina}%
). In order to construct the $(1+1)$-dimensional holographic superconductors
one should employ the $AdS_{3}/CFT_{2}$ correspondence. In \cite{chaturvedi}%
, the $(1+1)$-dimensional holographic superconductors were explored in the
probe limit and its distinctive features in both normal and superconducting
phases were investigated. Employing the variational method of the
Sturm-Liouville eigenvalue problem, the one-dimensional holographic
superconductors have been analytically studied in \cite{L12,momeni,peng17}.
It is also interesting to study the $(1+1)$-dimensional holographic
superconductor away from the probe limit by considering the backreaction. In
\cite{Wang}, the effects of backreaction have been studied for $s$-wave
linearly charged one-dimensional holographic superconductors.

Holographic superconductors have also been studied extensively in
the presence of nonlinear electrodynamics (see e.g.
\cite{n4,SH16,SSh16,SHsh(17),SHSH(16),shSh(16),n3,n5,n6}). The
most famous nonlinear electrodynamic is Born-Infeld
electrodynamic. This model was presented for the first time to
solve the problem of divergence of electrical field at the
position of point particle \cite{25,26,27,28,29}. It was later
showed that this model could be reproduced by string theory. In
the present work, we would like to extend the investigation on the $(1+1)$%
-dimensional holographic superconductors by taking into account
the nonlinear Born-Infeld (BI) electrodynamics, as our gauge
field. As well, we will study the effects of backreaction on our
holographic superconductors. We perform our investigation both
analytically and numerically and shall compare the result of two
methods. Our analytical approach is based on the Sturm-Liouville
variational method. In latter study, we find the relation between
critical temperature and chemical potential. Moreover, in order to
study our holographic superconductors numerically, we use the
shooting method. We show that analytical results are in good
agreement with numerical ones which implies that the
Sturm-Liouvile variation method is still powerful enough for
studying the $(1+1)$-dimensional holographic superconductor.

The structure of our paper is as follows. In section \ref{sec2}, the basic
field equations of one-dimensional holographic superconductors with
backreaction in the presence of BI nonlinear electrodynamics is introduced.
In section \ref{sec3}, we describe the procedure of analytical study of one
dimensional holographic superconductor based on Sturm-Liouvile method and
obtain the relation between critical temperature and chemical potential. In
section \ref{sec4}, the numerical approach toward the study of our
holographic superconductors will be presented. Finally, we summarize our
results in section \ref{sec5}.

\section{Basic field equations\label{sec2}}

The action of three dimensional AdS gravity in the presence of a gauge and a
scalar field is given by%
\begin{eqnarray}
S &=&\frac{1}{2\kappa ^{2}}\int d^{3}x\sqrt{-g}\left( R+\frac{2}{l^{2}}%
\right)  \notag \\
&&+\int d^{3}x\sqrt{-g}\left[ \mathcal{L}(\mathcal{F})-|\nabla \psi -iqA\psi
|^{2}-m^{2}|\psi |^{2}\right] ,  \notag \\
&&  \label{act}
\end{eqnarray}%
where $m$ and $q$ shows the mass and the charge of scalar field, $\kappa
^{2}={8\pi G_{3}}$ and $G_{3}$ is $3$-dimensional Newton gravitation
constant. Also, $g$, $R$ and $l$ are the metric determinant, Ricci scalar
and AdS radius, respectively. In (\ref{act}), $\mathcal{F}=F_{\mu \nu
}F^{\mu \nu }$ where $F_{\mu \nu }=\nabla _{\lbrack \mu }A_{\nu ]}$ is the
electrodynamics field tensor and $A_{\mu }$ is the vector potential. $%
\mathcal{L}(\mathcal{F})$ stands for the Lagrangian density of BI nonlinear
electrodynamics defined as%
\begin{equation}
\mathcal{L}(\mathcal{F})=\frac{1}{b}\left( 1-\sqrt{1+\frac{b\mathcal{F}}{2}}%
\right) ,
\end{equation}%
where $b$ is the nonlinear parameter. When $b\rightarrow 0$, $\mathcal{L}$
reduces to $-F_{\mu \nu }F^{\mu \nu }/{4}$ which is the standard Maxwell
Lagrangian \cite{H08}. Variation of the above action with respect to scalar
field $\psi $, gauge field $A_{\mu }$ and the metric $g_{\mu \nu }$ yield
the following equations of motion%
\begin{eqnarray}
0 &=&\left( \nabla _{\mu }-i{q}A_{\mu }\right) \left( \nabla ^{\mu }-i{q}%
A^{\mu }\right) \psi -m^{2}\psi \,,  \label{Epsi} \\
&&  \notag \\
0 &=&\nabla ^{\mu }\left( 4\mathcal{L}_{\mathcal{F}}F_{\mu \nu }\right)
\notag \\
&&-i{q}\left[ -\psi ^{\ast }(\nabla _{\nu }-i{q}A_{\nu })\psi +\psi (\nabla
_{\nu }+i{q}A_{\nu })\psi ^{\ast }\right] \,,  \label{02} \\
&&  \notag \\
0 &=&\frac{1}{2\kappa ^{2}}\left[ R_{\mu \nu }-g_{\mu \nu }\left( \frac{R}{2}%
+\frac{1}{l^{2}}\right) \right] +2F_{ac}F_{b}{}^{c}\mathcal{L}_{\mathcal{F}}
\notag \\
&&-\frac{g_{\mu \nu }}{2}\left[ \mathcal{L}(\mathcal{F})-m^{2}|\psi |^{2}-{%
|\nabla \psi -i{q}A\psi |^{2}}\right]  \notag \\
&&-\frac{1}{2}\left[ (\nabla _{\mu }\psi -i{q}A_{\mu }\psi )(\nabla _{\nu
}\psi ^{\ast }+i{q}A_{\nu }\psi ^{\ast })+\mu \leftrightarrow \nu \right] ,
\notag \\
&&  \label{Eein}
\end{eqnarray}%
where $\mathcal{L}_{\mathcal{F}}=\partial \mathcal{L}/\partial
{\mathcal{F}}$.

Since, we would like to consider the effect of the backreaction on the
holographic superconductor, we take a metric ansatz as follows \cite{Wang}%
\begin{equation}
{ds}^{2}=-f(r)e^{-\chi (r)}{dt}^{2}+\frac{{dr}^{2}}{f(r)}+\frac{r^{2}}{l^{2}}%
{dx}^{2}.  \label{metric}
\end{equation}%
The Hawking temperature of the three dimensional black hole on the
outer horizon $r_{+}$ (where $r_{+}$ is the horizon radius
obtained as the largest root of $f(r_{+})=0$), may be obtained
through the use of the general definition of surface gravity
\cite{SheyKaz}
\begin{eqnarray}\label{Tem1}
T&=&\frac{\kappa_{sg}}{2
\pi}=\frac{1}{2\pi}\sqrt{-\frac{1}{2}(\nabla_{\mu}\chi _{\nu})
(\nabla^{\mu}\chi^{\nu})},
\end{eqnarray}
where $\kappa_{sg}$ is the surface gravity and
$\chi=\partial/\partial t$ is the null killing vector of the
horizon. Taking $\chi^{\nu}=(-1,0,0)$, we have
$\chi_{\nu}=(f(r_{+}) e^{-\chi (r_{+})},0,0)$ and hence on the
horizon where $f(r_{+})=0$, we find $(\nabla_{\mu}\chi _{\nu})
(\nabla^{\mu}\chi^{\nu})=-\frac{1}{2}\left[f'(r_{+}) \right]^2
e^{-\chi (r_{+})}$. Thus, the temperature is obtained as
\begin{equation}
T=\frac{e^{-\chi (r_{+})/2}f^{^{\prime }}(r_{+})}{4\pi }.
\label{temp}
\end{equation}%
We also choose the scalar and the gauge fields as \cite{H08}%
\begin{equation}
A_{\mu }=(\phi (r),0,0),\ \ \ \psi =\psi (r).  \label{Aphi}
\end{equation}%
Substituting (\ref{metric}) and (\ref{Aphi}) into the field equations (\ref%
{Epsi})-(\ref{Eein}), we arrive at%
\begin{eqnarray}
0 &=&\psi ^{\prime \prime }+\psi ^{\prime }\left[ \frac{1}{r}+\frac{%
f^{\prime }}{f}-\frac{\chi ^{\prime }}{2}\right] +\psi \left[ \frac{%
q^{2}\phi ^{2}{\mathrm{e}}^{\chi }}{f^{2}}-\frac{m^{2}}{f}\right] ,
\label{psir} \\
&&  \notag \\
0 &=&\phi ^{\prime \prime }+\phi ^{\prime }\left[ -\frac{be^{\chi }\phi
^{\prime 2}}{r}+\frac{\chi ^{\prime }}{2}+\frac{1}{r}\right] -\frac{%
2q^{2}\psi ^{2}\phi }{f}\left[ 1-be^{\chi }\phi ^{\prime 2}\right] ^{3/2},
\notag \\
&&  \label{phir} \\
0 &=&f^{\prime }-\frac{2r}{l^{2}}  \notag \\
&&+2\kappa ^{2}r\left[ \frac{q^{2}e^{\chi }\psi ^{2}\phi ^{2}}{f}+f\psi
^{\prime 2}+m^{2}\psi ^{2}-\frac{1}{b}+\frac{1}{b\sqrt{1-be^{\chi }\phi
^{\prime 2}}}\right] ,  \notag \\
&&  \label{fr} \\
0 &=&\chi ^{\prime }+4\kappa ^{2}r\left[ \frac{q^{2}\phi ^{2}\psi ^{2}{%
\mathrm{e}}^{\chi }}{f^{2}}+\psi ^{\prime 2}\right] ,  \label{chir}
\end{eqnarray}%
where the prime denotes derivative with respect to $r$. Note that in the
presence of nonlinear BI electrodynamics the Eqs. (\ref{psir}) and (\ref%
{chir}) do not change compared to the linear Maxwell case. In the limiting
case where $b\rightarrow 0$ the equations of motion (\ref{phir}) and (\ref%
{fr}) turn to the corresponding equations of one dimensional holographic
superconductor with Maxwell field \cite{Wang}. The field equations (\ref%
{psir})-(\ref{chir}) enjoy the symmetries
\begin{gather}
q\rightarrow q/a,\text{ \ \ \ \ }\phi \rightarrow a\phi ,\text{ \ \ \ \ }%
\psi \rightarrow a\psi ,  \notag \\
\kappa \rightarrow \kappa /a,\text{ \ \ \ \ }b\rightarrow b/a^{2}, \\
\notag \\
l\rightarrow al,\text{ \ \ \ \ }r\rightarrow ar,\text{ \ \ \ \ }q\rightarrow
q/a,  \notag \\
m\rightarrow m/a,\text{ \ \ \ \ }b\rightarrow a^{2}b.
\end{gather}%
Hereafter, we set $q$ and $l$ equal to unity by virtue of these symmetries.
The behavior of our model functions for large $r$ (near the boundary) read%
\begin{gather}
\chi \rightarrow 0,\ \ \ f(r)\sim r^{2},  \notag \\
\phi \sim \rho +\mu \ln (r),\ \ \ \psi \sim \psi _{-}r^{-\Delta _{-}}+\psi
_{+}r^{-\Delta _{+}},  \label{bval}
\end{gather}%
where $\mu $ and $\rho $ are the chemical potential and the charge
density of the field theory on the boundary and $\Delta _{\pm
}=1\pm \sqrt{1+m^{2}}$ which implies $m^{2}\geq -1$. Actually,
$\chi $ could be a constant near the boundary but by using the
symmetry of field equation $\mathrm{e}^{\chi }\rightarrow
a^{2}\mathrm{e}^{\chi },$ $\phi \rightarrow \phi /a$, it could be
set to zero there. From holographic superconductors point of view,
either $\psi _{+}$ or $\psi _{-}$ can be dual to the expectation
value of condensation operator (or order operator) $\left\langle
O\right\rangle $ while the other is dual to its source. We give
$\psi _{-}$ the role of
source and $\psi _{+}$ the role of expectation value of the order parameter $%
\left\langle O_{+}\right\rangle $ in this work. Since we seek for study the
effects of $b$ and $\kappa $ on our holographic superconductors and
different values of the scalar field mass do not influence this behavior
qualitatively, we consider $m^{2}=0$ in this work. With this choice, we have
$\Delta _{+}=2$, $\Delta _{-}=0$ and thus%
\begin{equation}\label{psiasy}
\psi \sim \psi _{-}+\frac{\psi _{+}}{r^{2}},
\end{equation}%
near the boundary. We set $\psi _{-}=0$ at the boundary and
consider $\psi _{+}$ as the dual of order parameter $\left\langle
O_{+}\right\rangle $. It is remarkable to note that the asymptotic
solution for $\psi$ given in Eq. (\ref{psiasy}) do not depend on
the type of electrodynamics and thus for the Maxwell case in three
dimensions the solution is the same as in  Eq. (\ref{psiasy}).
While the solution depends on the spacetime dimensions. This is
due to the fact that equation for the $\psi$  given in
(\ref{psir}) is independent on the type of electrodynamics but
depends on the spacetime dimensions and the mass parameter $m$
\cite{Wang,Doa,ghor}.

The next step is to solve the coupled nonlinear field equations (\ref{psir}%
)-(\ref{chir}) simultaneously and obtain the behavior of model functions.
Then, we could figure out the behavior of different parameters of
holographic superconductor specially the order parameter $\left\langle
O_{+}\right\rangle $ and the critical temperature by using these functions.
In this work, we use both analytical and numerical methods for studying the
holographic superconductor. For analytical study, we perform Sturm-Liouville
method while for numerical study, we use shooting method.

\section{Sturm-Liouville method\label{sec3}}

\begin{table*}[t]
\caption{Analytical results of ${T_{c}}/{\protect\mu }$ for different values
of backreaction and nonlinearity parameters.}
\label{tab1}
\begin{center}
\begin{tabular}{c|c|c|c|c|c|c|}
\cline{2-3}\cline{2-7}\cline{4-7}
& \multicolumn{2}{|c|}{$b=0$} & \multicolumn{2}{|c|}{$b=0.04$} &
\multicolumn{2}{|c|}{$b=0.08$} \\ \cline{2-3}\cline{2-7}\cline{4-7}
& Analytical & Numerical & Analytical & Numerical & Analytical & Numerical
\\ \hline
\multicolumn{1}{|c|}{$\kappa ^{2}=0$} & $0.0429$ & $0.0460$ & $0.0360$ & $%
0.0410$ & $0.0275$ & $0.0362$ \\ \hline
\multicolumn{1}{|c|}{$\kappa ^{2}=0.05$} & $0.0399$ & $0.0369$ & $0.0337$ & $%
0.0326$ & $0.0260$ & $0.0286$ \\ \hline
\multicolumn{1}{|c|}{$\kappa ^{2}=0.1$} & $0.0381$ & $0.0295$ & $0.0311$ & $%
0.0260$ & $0.0218$ & $0.0227$ \\ \hline
\multicolumn{1}{|c|}{$\kappa ^{2}=0.15$} & $0.0352$ & $0.0236$ & $0.0280$ & $%
0.0207$ & $0.0174$ & $0.0180$ \\ \hline
\multicolumn{1}{|c|}{$\kappa ^{2}=0.2$} & $0.0313$ & $0.0189$ & $0.0242$ & $%
0.0165$ & $0.0136$ & $0.0143$ \\ \hline
\multicolumn{1}{|c|}{$\kappa ^{2}=0.25$} & $0.0264$ & $0.0151$ & $0.0195$ & $%
0.0131$ & $0.0089$ & $0.0114$ \\ \hline
\end{tabular}%
\end{center}
\end{table*}

In this section, we employ the Sturm-Liouville eigenvalue problem to
investigate analytically the phase transition of one dimensional $s$-wave
holographic superconductor in the presence of BI nonlinear electrodynamics.
In addition, we calculate the relation between the critical temperature $%
T_{c}$, and chemical potential $\mu $, near the horizon. Furthermore, we
study the effect of backreaction and BI nonlinear electrodynamics on the
critical temperature. For future convenience, we define a new variable $%
z=r_{+}/r$ $(\in \left[ 0,1\right] )$. With this new coordinate, the field
equations (\ref{psir})-(\ref{chir}) could be rewritten as%
\begin{eqnarray}
0 &=&\psi ^{\prime \prime }+\psi ^{\prime }\left[ \frac{f^{\prime }}{f}-%
\frac{\chi ^{\prime }}{2}+\frac{1}{z}\right] +\psi \left[ \frac{%
r_{+}^{2}e^{\chi }\phi ^{2}}{z^{4}f^{2}}-\frac{m^{2}r_{+}^{2}}{z^{4}f}\right]
,  \notag \\
&&  \label{psiz} \\
0 &=&\phi ^{\prime \prime }+\phi ^{\prime }\left[ \frac{bz^{3}e^{\chi }\phi
^{\prime 2}}{r_{+}^{2}}+\frac{\chi ^{\prime }}{2}+\frac{1}{z}\right] -\frac{%
2r_{+}^{2}\psi ^{2}\phi }{z^{4}f}\Upsilon ^{\frac{3}{2}},  \label{phiz} \\
&&  \notag \\
0 &=&f^{\prime }+\frac{2r_{+}^{2}}{z^{3}}+\frac{2r_{+}^{2}\kappa ^{2}}{z^{3}}
\notag \\
&&\times \left[ \frac{1}{b}\left( 1-\Upsilon ^{-\frac{1}{2}}\right) -\frac{%
z^{4}f\psi ^{\prime 2}}{r_{+}^{2}}-\frac{e^{\chi }\psi ^{2}\phi ^{2}}{f}%
-m^{2}\psi ^{2}\right] ,  \notag \\
&&  \label{fz} \\
0 &=&\chi ^{\prime }-4\kappa ^{2}\left[ \frac{r_{+}^{2}e^{\chi }\psi
^{2}\phi ^{2}}{z^{3}f^{2}}+z\psi ^{\prime 2}\right] ,  \label{chiz}
\end{eqnarray}%
where $\Upsilon =1-bz^{4}e^{\chi }\phi ^{\prime 2}/r_{+}^{2}$ and now the
prime indicates the derivative with respect to $z$. Since in the vicinity of
critical temperature the order parameter is small, we can consider it as an
expansion parameter%
\begin{equation*}
\epsilon \equiv \left\langle O_{i}\right\rangle ,
\end{equation*}%
where $i=+$ or $-$. We focus on solutions for small values of condensation
parameter $\epsilon $, therefore we can expand the model functions as%
\begin{gather*}
\psi \approx \epsilon \psi _{1}+\epsilon ^{3}\psi _{3}+\epsilon ^{5}\psi
_{5}+\cdots , \\
\phi \approx \phi _{0}+\epsilon ^{2}\phi _{2}+\epsilon ^{4}\phi _{4}+\cdots ,
\\
f\approx f_{0}+\epsilon ^{2}f_{2}+\epsilon ^{4}f_{4}+\cdots , \\
\chi \approx \epsilon ^{2}\chi _{2}+\epsilon ^{4}\chi _{4}+\cdots ,
\end{gather*}%
where $\epsilon \ll 1$ near the critical temperature. Moreover, by
considering $\delta \mu _{2}>0$, the chemical potential can be expressed as:%
\begin{equation*}
\mu =\mu _{0}+\epsilon ^{2}\delta \mu _{2}+...\rightarrow \epsilon
\thickapprox \Bigg(\frac{\mu -\mu _{0}}{\delta \mu _{2}}\Bigg)^{1/2}.
\end{equation*}%
During phase transition, $\mu _{c}=\mu _{0}$, thus the order parameter
vanishes. Meanwhile, the critical exponent $\beta =\frac{1}{2}$ is in a good
agreement with mean field theory result.

At zeroth order of $\epsilon $, the gauge field equation of motion (\ref%
{phiz}) reduces to%
\begin{equation}
\phi ^{\prime \prime }+\frac{\phi ^{\prime }}{z}+\frac{bz^{3}\phi ^{\prime 3}%
}{r_{+}^{2}}=0,
\end{equation}%
which could be rewritten as a first order Bernoulli differential equation by
taking $\phi ^{\prime }$ as a new function \cite{TL00}. Therefore, one
receives%
\begin{equation}
\phi ^{\prime }=\frac{\lambda r_{+}}{z\sqrt{b\lambda ^{2}z^{2}+1}},
\label{dphi}
\end{equation}%
for small values of $b$ where we define $\lambda =\mu /r_{+}$ and fix the
integration constants by looking at the behavior of $\phi $ near the
boundary given in (\ref{bval}). Integrating (\ref{dphi}) and using the fact
that $\phi (z=1)=0$\footnote{%
It is necessary so that the norm of gauge potential is finite at horizon.},
we can obtain%
\begin{eqnarray}
\phi _{0}(z) &=&\int_{1}^{z}\frac{\lambda r_{+}}{z}\left( 1-\frac{1}{2}%
b\lambda ^{2}z^{2}\right) \,dz  \notag \\
&=&\lambda r_{+}\log (z)-\frac{1}{4}b\lambda ^{3}r_{+}\left( z^{2}-1\right) .
\label{phi}
\end{eqnarray}%
When $b=0$ the above equation reduces to one of \cite{L12}. Note that at the
zeroth order with respect to $\epsilon $, $\psi _{0}=\chi _{0}=0$.
Substituting (\ref{dphi}) in the (\ref{fz}), the equation for $f$ at the
zeroth order with respect to $\epsilon $ has the following form

\begin{gather}
f_{0}(z)=r_{+}^{2}g(z),  \notag \\
\text{\ }g(z)=\frac{1}{z^{2}}-1+\frac{1}{8}b\kappa ^{2}\lambda ^{4}\left(
1-z^{2}\right) +\kappa ^{2}\lambda ^{2}\log (z).
\end{gather}

\begin{figure*}[t]
\centering
\subfigure[~b=0]{\includegraphics[width=0.3\textwidth]{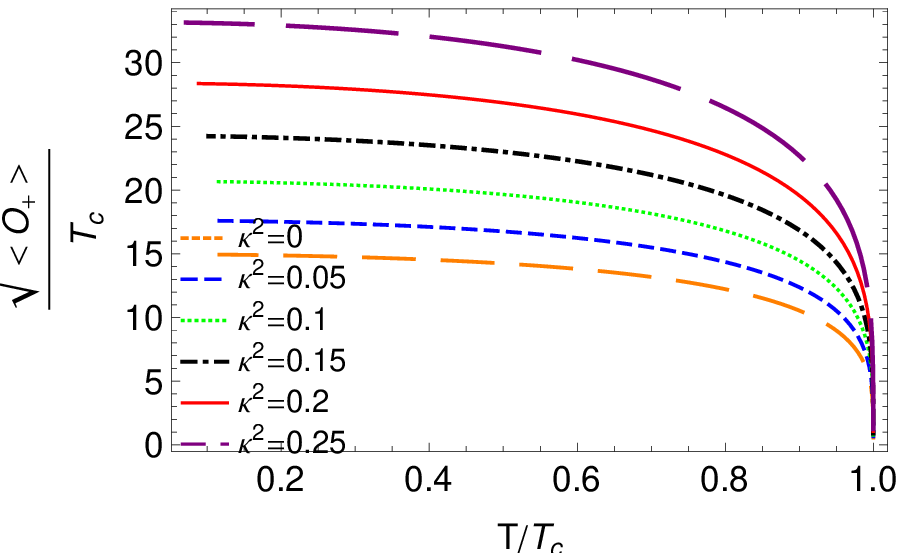}} \qquad %
\subfigure[~b=0.04]{\includegraphics[width=0.3\textwidth]{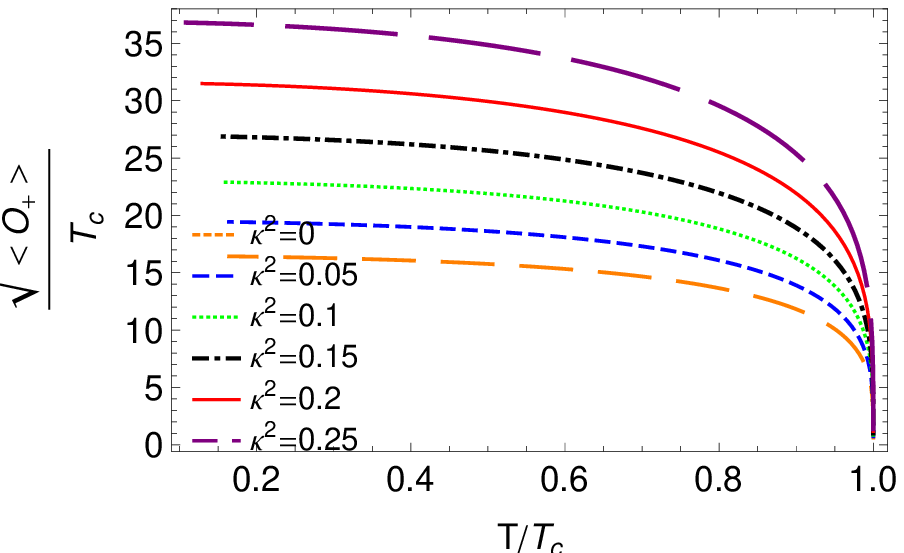}} \qquad %
\subfigure[~b=0.08]{\includegraphics[width=0.3\textwidth]{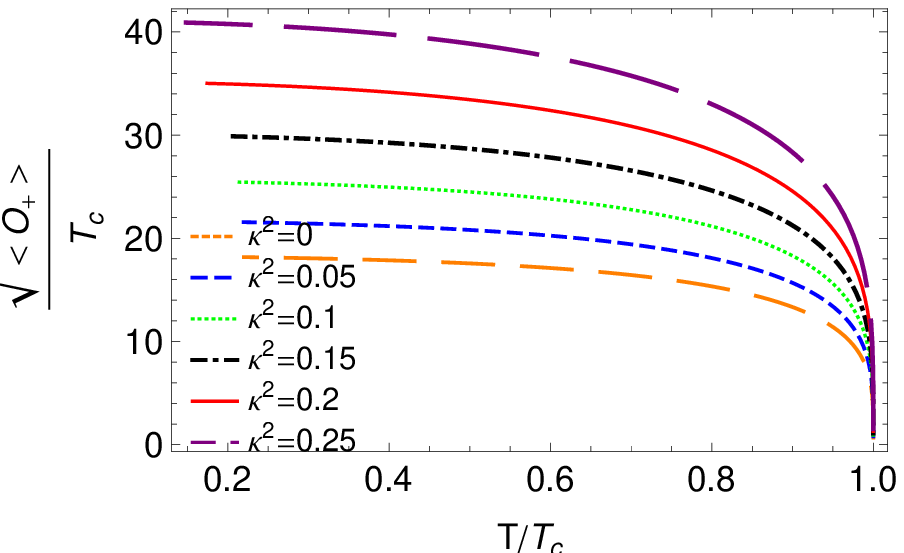}}
\caption{The behavior of condensation parameter as a function of temperature
for different values of backreaction.}
\label{fig1}
\end{figure*}

\begin{figure*}[t]
\centering
\subfigure[~$\kappa^{2}$=0]{\includegraphics[width=0.3\textwidth]{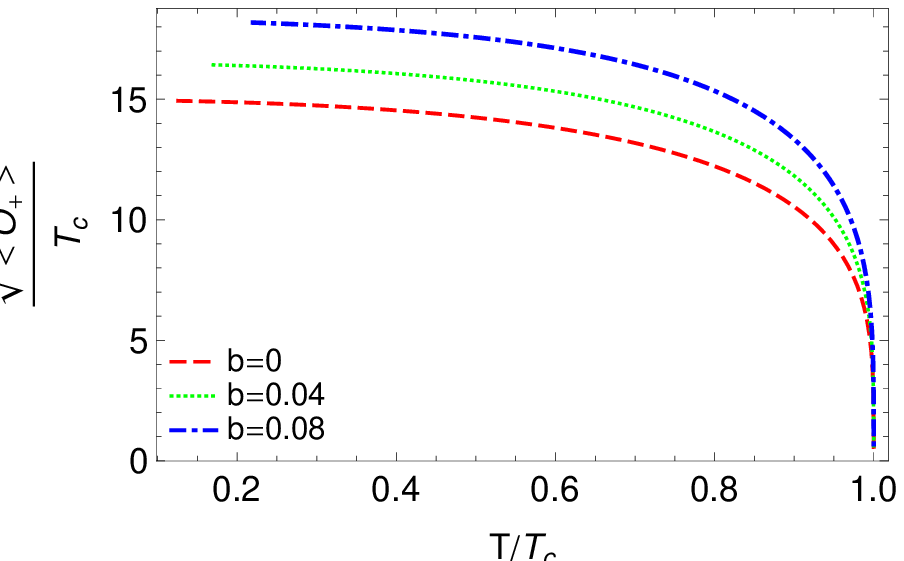}}
\qquad %
\subfigure[~$\kappa^{2}$=0.10]{\includegraphics[width=0.3%
\textwidth]{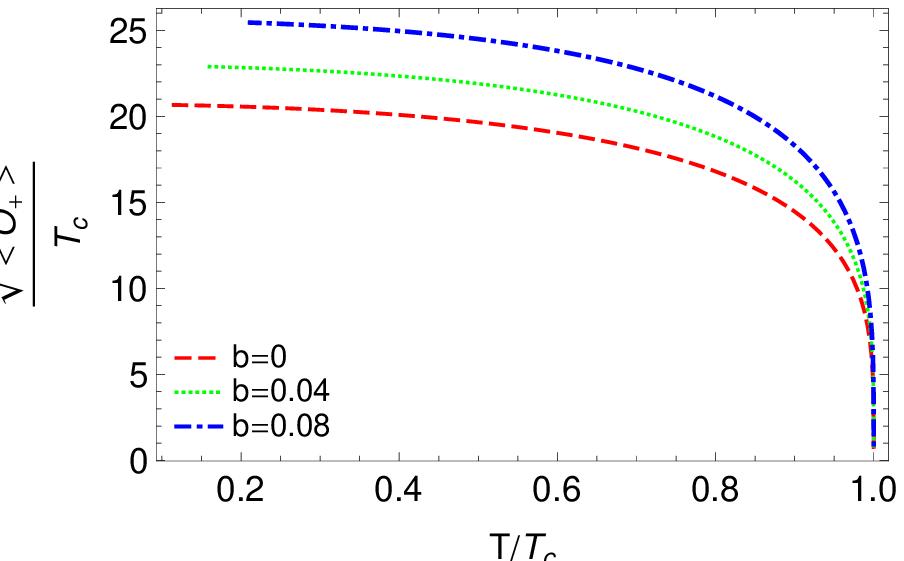}} \qquad %
\subfigure[~$\kappa^{2}$=0.20]{\includegraphics[width=0.3%
\textwidth]{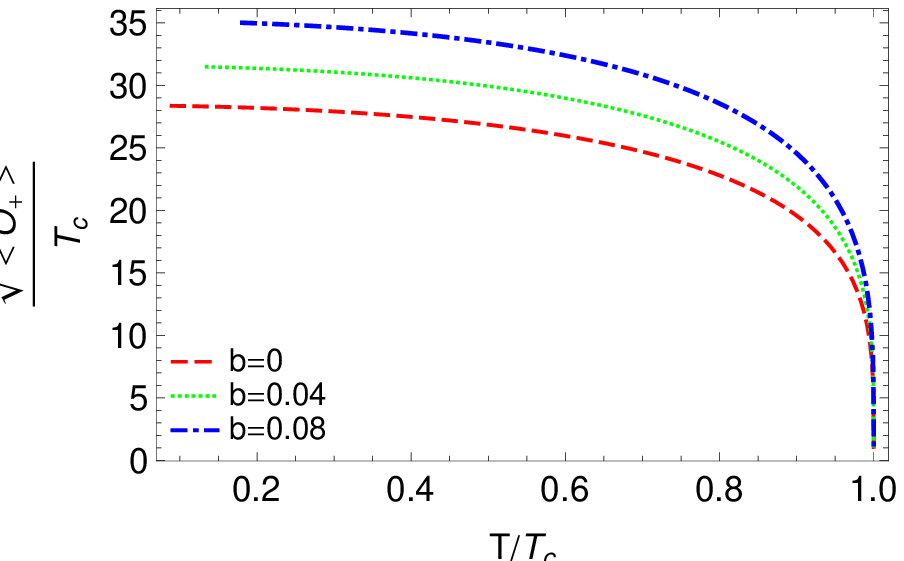}} \qquad %
\caption{The behavior of condensation parameter as a function of temperature
for different values of nonlinearity parameter $b$.}
\label{fig2}
\end{figure*}

The asymptotic behavior of the scalar field $\psi $ was given in (\ref{bval}%
). Near the boundary, we define the function $F(z)$ so that
\begin{equation}
\psi (z)=\frac{\left\langle O_{i}\right\rangle z^{\triangle _{i}}}{\sqrt{2}%
r_{+}^{\triangle _{i}}}F(z).  \label{psi1F}
\end{equation}%
Inserting Eq. (\ref{psi1F}) in Eq. (\ref{psiz}) yields%
\begin{eqnarray}
&&F^{\prime \prime }(z)+F^{\prime }(z)\left[ \frac{g^{\prime }(z)}{g(z)}+%
\frac{2\Delta }{z}+\frac{1}{z}\right]   \notag \\
&+&F(z)\left[ \frac{\Delta g^{\prime }(z)}{zg(z)}-\frac{m^{2}}{z^{4}g(z)}+%
\frac{\Delta ^{2}}{z^{2}}\right]   \notag \\
&-&\frac{F(z)}{2z^{4}g(z)^{2}}\left[ \lambda ^{2}\log (z)\left( b\lambda
^{2}r_{+}\left( z^{2}-1\right) -2\log (z)\right) \right] =0.  \notag \\
&&
\end{eqnarray}%
We can rewrite this equation in the Sturm-Liouville form as%
\begin{equation}
\left[ T(z)F^{\prime }(z)\right] ^{\prime }+P(z)T(z)F(z)+\lambda
^{2}Q(z)T(z)F(z)=0,  \label{sl}
\end{equation}%
where the functions $T$, $P$, $Q$ are defined as%
\begin{equation}
T(z)=z^{2\Delta +1}\left[ \frac{1}{z^{2}}-1+\frac{1}{8}b\kappa ^{2}\lambda
^{4}\left( 1-z^{2}\right) +\kappa ^{2}\lambda ^{2}\log (z)\right] .
\end{equation}%
\begin{equation}
P(z)=\frac{\Delta }{z}\left( \frac{g^{\prime }(z)}{g(z)}+\frac{\Delta }{z}%
\right) -\frac{m^{2}}{z^{4}g(z)},
\end{equation}%
\begin{equation}
Q(z)=-\frac{\log (z)\left( b\lambda ^{2}r_{+}\left( z^{2}-1\right) -2\log
(z)\right) }{2z^{4}g(z)^{2}}.
\end{equation}%
We can consider the trial function $F(z)=1-\alpha z^{2}$ which satisfies the
required boundary conditions $F(0)=1$ and $F^{^{\prime }}(0)=0$. Then, the
eigenvalue problem could be solved for (\ref{sl}) by minimizing the
expression%
\begin{equation}
\lambda ^{2}=\frac{\int_{0}^{1}T\left( F^{\prime 2}-PF^{2}\right) dz}{%
\int_{0}^{1}TQF^{2}dz},  \label{l2}
\end{equation}%
with respect to $\alpha $. For backreacting parameter, we could use the
iteration method and define \cite{LPJW2015}
\begin{equation}
\kappa _{n}=n\Delta \kappa ,\ \ \ n=0,1,2,\cdots ,
\end{equation}%
where $\Delta \kappa =\kappa _{n+1}-\kappa _{n}$. Here, we take $\Delta
\kappa =0.05$. Since we are interested in finding the effects of
nonlinearity on backreaction up to the order $\kappa ^{2}$, we have%
\begin{equation}
\kappa ^{2}\lambda ^{2}={\kappa _{n}}^{2}\lambda ^{2}={\kappa _{n}}%
^{2}(\lambda ^{2}|_{\kappa _{n-1}})+O[(\Delta \kappa )^{4}],
\end{equation}%
where we take $\kappa _{-1}=0$ and $\lambda ^{2}|_{\kappa _{-1}}=0$. We
shall also retain the linear terms with respect to nonlinearity parameter $b$
and therefore,%
\begin{equation}
b\lambda ^{2}=b\left( \lambda ^{2}|_{b=0}\right) +\mathcal{O}(b^{2}).
\end{equation}%
Then, the minimum eigenvalue of Eq. (\ref{l2}) can be obtained. At the
critical point, temperature is defined as (see Eq. (\ref{temp}) and note
that at zeroth order with respect to $\epsilon $, $\chi $ is zero.)%
\begin{equation}
T_{c}=\frac{f^{\prime }\left( r_{+c}\right) }{4\pi }.
\end{equation}%
Using Eqs. (\ref{fr}) and (\ref{phi}), we receive
\begin{equation}
f^{\prime }\left( r_{+c}\right) =2r_{+c}+\frac{2\kappa ^{2}r_{+c}}{b}\left[
1-\frac{1}{\sqrt{1-b\phi ^{\prime }\left( r_{+c}\right) {}^{2}}}\right] ,
\label{frc}
\end{equation}%
and thus

\begin{eqnarray}
T_{c} &=&\frac{1}{4\pi }(\frac{\mu }{\lambda })[2-\kappa _{n}^{2}(\lambda
^{2}|_{\kappa _{n-1}})  \notag \\
&&-\frac{3}{4}b\kappa _{n}^{2}(\lambda ^{4}|_{\kappa _{n-1},b=0})+b\kappa
_{n}^{2}(\lambda ^{4}|_{\kappa _{n-1},b=0})].
\end{eqnarray}%
As an example, if $b=\kappa ^{2}=0$ we have%
\begin{equation*}
\lambda ^{2}=\dfrac{\frac{2}{3}\alpha ^{2}-\frac{4}{3}\alpha +1}{-\frac{%
251\alpha ^{2}}{864}+\frac{9\alpha }{16}+\frac{\alpha ^{2}\zeta (3)}{4}-%
\frac{\alpha \zeta (3)}{2}+\frac{\zeta (3)}{4}-\frac{1}{4}}.
\end{equation*}%
Inserting $\alpha =0.759$, $\lambda _{min}^{2}=13.76$ and $T_{c}=0.429\mu $.
The latter result perfectly agrees with ones in \cite{L12}.

The values of $T_{c}/\mu $ for different backreaction and nonlinearity
parameters are listed in \ref{tab1}. As it shows, the effect of increasing
the backreaction parameter $\kappa $ for a fixed value of nonlinearity
parameter $b$ follows the same trend as raising $b$ for a fixed value of $%
\kappa $. In both cases, the critical temperature $T_{c}$ diminishes by
growing the backreaction or nonlinearity parameters. It shows that the
presence of backreaction and Born-Infeld nonlinear electrodynamics make the
scalar hair harder to form. In next section, we will re-study the problem
numerically using the shooting method.

\section{Shooting method\label{sec4}}

In this section, we will study our holographic superconductor numerically.
In order to do this, we use the shooting method \cite{H09}. In this method,
the boundary values is found by setting appropriate initial conditions. So,
for doing this, we need to know the behavior of equations of motion both at
horizon and boundary. Using Taylor expansion at horizon around $z=1$, we get%
\begin{gather}
f(z)=f_{1}\left( 1-z\right) +f_{2}\left( 1-z\right) {}^{2}+\cdots , \\
\phi (z)=\phi _{1}\left( 1-z\right) +\phi _{2}\left( 1-z\right)
{}^{2}+\cdots , \\
\psi (z)=\psi _{0}+\psi _{1}\left( 1-z\right) +\psi _{2}\left( 1-z\right)
{}^{2}+\cdots , \\
\chi (z)=\chi _{0}+\chi _{1}\left( 1-z\right) +\chi _{2}\left( 1-z\right)
{}^{2}+\cdots .
\end{gather}%
Note that $\phi =0$ at horizon, otherwise it will be ill-defined there. In
our procedure, we find all coefficients in terms of $\phi _{1}$, $\psi _{0}$
and $\chi _{0}$ by using equations of motion. Varying them at the horizon,
we try to get $\psi _{-}=\chi =0$ at the boundary. So, the values of $\psi
_{+}$ and $\mu $ are achieved. In addition, we will set $r_{+}=1$ by virtue
of the equations of motion's symmetry%
\begin{equation*}
r\rightarrow ar,\text{ \ \ \ \ }f\rightarrow a^{2}f,\text{ \ \ \ \ }\phi
\rightarrow a\phi .
\end{equation*}

Performing numerical solution, we can find the values of $T_{c}/\mu $ for
different backreaction and nonlinearity parameters. In order to compare the
latter results with analytical ones, we listed both of them next to each
other in table \ref{tab1}. It is obvious that there is a reasonable
agreement between the results of both methods. Moreover, in table \ref{tab1}%
, the results of \cite{Wang} for $b=0$ has been recovered for different
values of backreaction parameter. As one could see in this table, increasing
the backreaction parameter for a fixed value of $b$, decreases the critical
temperature. This means that the larger values of backreaction parameter
makes the condensation harder to form. Similarly, for a fixed value of $%
\kappa $, increasing the nonlinearity of electrodynamic model makes scalar
hair harder to form because it diminishes the critical temperature.

Figs. \ref{fig1} and \ref{fig2} give information about the effect of
backreaction and nonlinear electrodynamic on condensation. All curves follow
a same trend. As $b\rightarrow 0$, we regain the results of Maxwell case
presented in \cite{Wang}. As figures show, the condensation gap increases by
making backreaction and nonlinearity parameters larger while the other one
is fixed. So, it can be understood that it is harder to form a
superconductor. This is in agreement with the results obtained from the
behavior of critical temperature before.

\section{Summary and discussion\label{sec5}}

In this work, by using the Sturm-Liouville eigenvalue problem, we
analytically investigated the properties of $(1+1)$-dimensional holographic
superconductor developed in BTZ black hole background in the presence of BI
nonlinear electrodynamics. We have relaxed the probe limit and further
assumed that the gauge and scalar fields do backreact on the background
metric. We determined the critical temperature for different values of
backreaction and nonlinear parameters. We have also continued our study by
using the numerical shooting method and confirmed that the analytical
results are in agreement with the numerical approach. We observed that the
formation of the scalar hair is harder in the presence of BI nonlinear
electrodynamics as well as backreaction and it becomes harder and harder to
form by increasing the strength of either the nonlinear and backreaction
parameters.

Finally, it would be of interest to extend this procedure for other
nonlinear electrodynamics like Power-Maxwell and logarithmic cases and
investigate the effects of nonlinear electrodynamics on the critical
temperature and condensation operator of one dimensional holographic
superconductors. These issues are now under investigations and the results
will be appeared elsewhere.


\begin{acknowledgments}
We thank referee for constructive comments which helped us improve
our paper significantly. We also thank Shiraz University Research
Council. The work of AS has been supported financially by Research
Institute for Astronomy and Astrophysics of Maragha (RIAAM), Iran.
MKZ would like to thank Shahid Chamran University of Ahvaz, Iran.
\end{acknowledgments}

\end{document}